\documentclass[twocolumn,aps,prl,amssymb,showpacs,superscriptaddress]{revtex4}

\usepackage{graphicx}
\usepackage{dcolumn}
\usepackage{bm}

\begin{document}
\preprint{PREPRINT (\today)}

\title{A metallic phase in lightly doped La$_{2-x}$Sr$_{x}$CuO$_{4}$ observed by 
electron paramagnetic resonance}
\author{A.~Shengelaya} 
\affiliation{Physik-Institut der Universit\"{a}t
Z\"{u}rich, Winterthurerstrasse 190, CH-8057, Switzerland}
\author{M.~Bruun} 
\affiliation{Physik-Institut der Universit\"{a}t
Z\"{u}rich, Winterthurerstrasse 190, CH-8057, Switzerland}
\author{B.I.~Kochelaev} 
\affiliation{Department of Physics, Kazan State University, Kazan, 420008, 
Russia}
\author{A.~Safina} 
\affiliation{Department of Physics, Kazan State University, Kazan, 420008, 
Russia}
\author{K.~Conder}
\affiliation{Laboratory for Neutron Scattering, ETH Z\"urich and
PSI, CH-5232 Villigen PSI, Switzerland}
\author{K.A.~M\"uller} 
\affiliation{Physik-Institut der Universit\"{a}t Z\"{u}rich,
Winterthurerstrasse 190, CH-8057, Switzerland}
%

\begin{abstract}

In the low doping range of $x$ from 0.01 to 0.06 in
La$_{2-x}$Sr$_{x}$CuO$_{4}$, a narrow electron paramagnetic resonance
(EPR) line has been investigated.  This line is distinct from the
known broad line, both due to probing Mn$^{2+}$ ions.  The narrow line
is ascribed to metallic regions in the material, and its intensity
increases exponentially upon cooling below $\sim$ 150 K. The
activation energy deduced $\Delta$ = 460(50) K is nearly the same as
that found in the doped superconducting regime by Raman and neutron
scattering.  The intensity of the narrow EPR line follows the same
temperature dependence as the resistivity anisotropy in lightly doped
La$_{2-x}$Sr$_{x}$CuO$_{4}$ single crystals.

\end{abstract}

\maketitle

The generic phase diagram in hole-doped cuprates is by now well
established.  At a critical concentration of doping $x_{c1}\approx$ 
0.06, superconductivity sets in at $T$ = 0, and ends at a higher doping 
level $x_{c2}\approx$ 0.25 \cite{Tallon}.  
Both are the critical endpoints of the superconducting phase-
transition line \cite{Schneider}.  At the former, a transition from an insulating to 
the superconducting state has been assumed  untill very recently \cite{Schneider}.  
However, using finite-size scaling for the susceptibility
of a series of concentrations $x<x_{c1}$, the following was inferred: The
material consists of antiferromagnetic (AF) domains of variable size,
separated by metallic domain walls \cite{Johnston}.  More recently 
Ando {\em et al.} corroborated this early finding by measuring the in-plane
resistivity anisotropy in untwinned single crystals of
La$_{2-x}$Sr$_{x}$CuO$_{4}$ (LSCO) and YBa$_{2}$Cu$_{3}$O$_{7-\delta}$
(YBCO) in the lightly doped region, interpreting their results in
terms of metallic stripes present \cite{Ando1}.  
Previouse electron paramagnetic resonance (EPR)  
measurements in LSCO at $x = 0.03<x_{c1}$ revealed in addition 
to a broad EPR line observed in the entire doping range $x$ \cite{Alika1}, 
a narrow line \cite{Alika2}.  It was interpreted as stemming from a metallic phase
distinct from the volume in which the broader line is due to.  The latter
showed a clear isotope effect in its linewidth \cite{Alika1,Alika2}, the narrow
one did not.  In the present letter, we describe a thorough EPR
investigation of the behavior of the narrow line for concentrations 
0.01 $\leq x\leq$ 0.06, i.e.  below $x_{c1}$.  Of special interest is the
exponential  increase of the narrow-line intensity upon cooling.  The
activation energy inferred is nearly the same as that deduced from
other experiments for the formation of bipolarons \cite{Bipolaron}, pointing
to the origin of the metallic stripes present.

\begin{figure}[htb]
\includegraphics[width=0.8\linewidth]{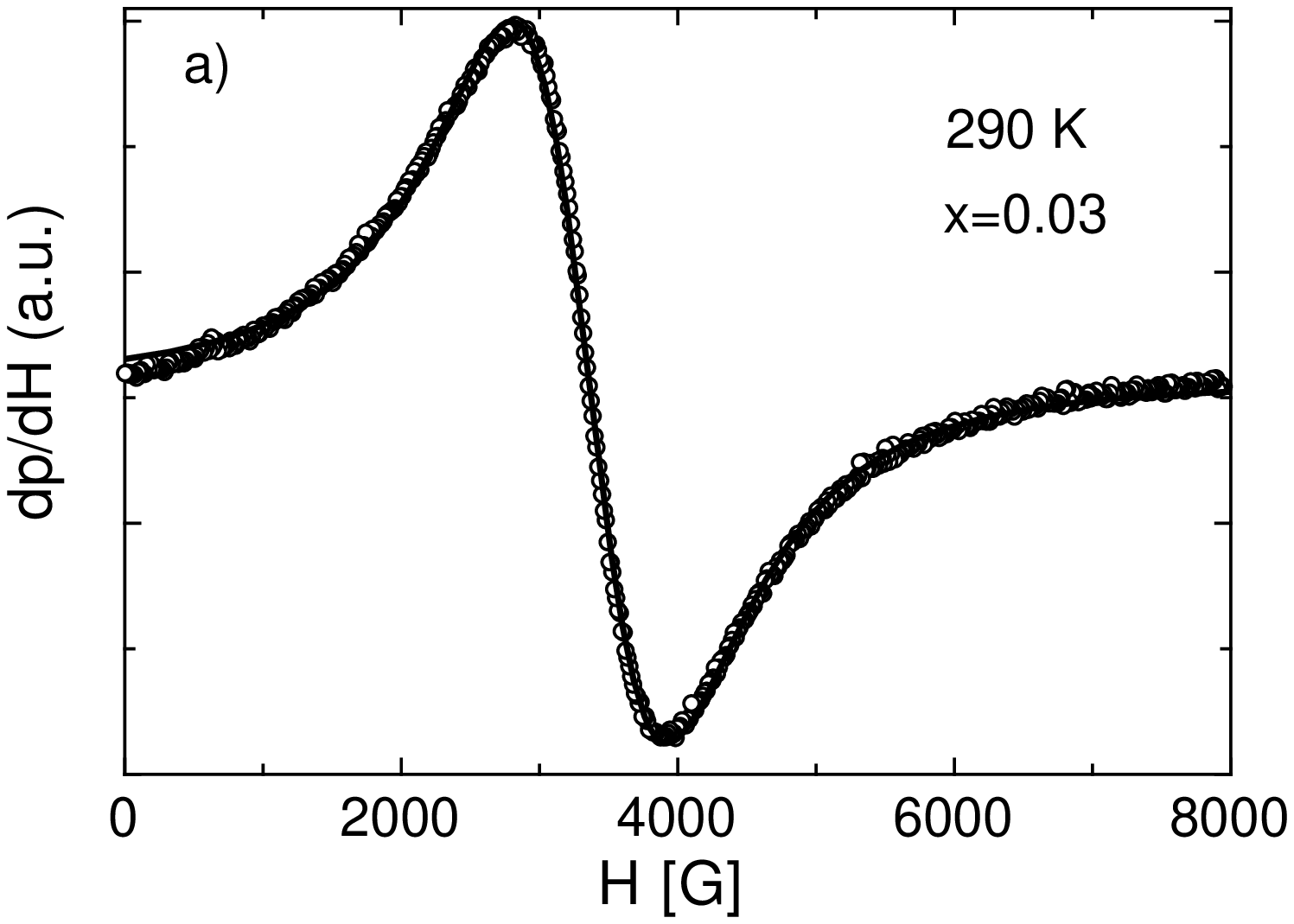}
\includegraphics[width=0.8\linewidth]{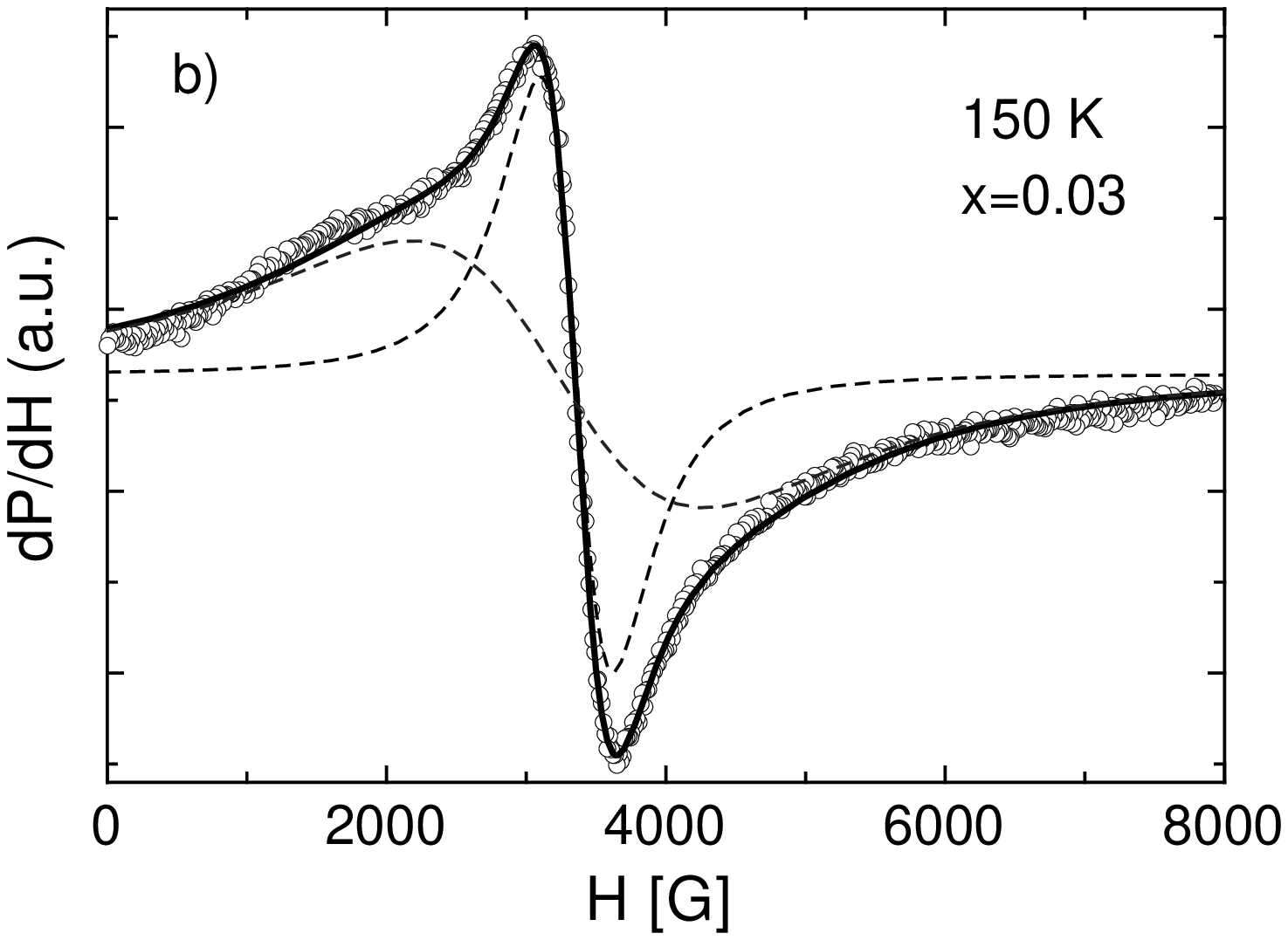}
\caption{(a) EPR signal of 
La$_{1.97}$Sr$_{0.03}$Cu$_{0.98}$Mn$_{0.02}$O$_{4}$
measured at $T$ = 290 K. 
The fit to a Lorentzian line shape is represented by the solid line.  
(b) EPR signal at $T$=150 K. The solid line is a fit with a sum of two
Lorentzians represented by dashed lines.}
\label{Fig.1}
\end{figure}

The La$_{2-x}$Sr$_{x}$Cu$_{0.98}$Mn$_{0.02}$O$_{4}$ polycrystalline samples 
with 0$\leq$x$\leq$0.06  were prepared by the standard solid-state reaction method.  
The EPR measurements were performed at 9.4 GHz using a BRUKER ER-200D 
spectrometer equipped with an Oxford Instruments helium flow 
cryostat.  In order to avoid a signal distortion due to skin effects, the samples 
were ground and the powder was suspended in paraffin.  
We observed an EPR signal in all samples.  The signal is centered near
$g\sim 2$, a value very close to the $g$-factor for the Mn$^{2+}$ ion.  
Figure 1 shows typical EPR spectra for an $x$ = 0.03 sample at two different 
temperatures. One can see that at 290 K only a single EPR line of Lorentzian shape 
is observed.  However, with decreasing temperature a second line appears 
(see Fig.  1 (b)), and the EPR spectra can be well fitted by a sum of two Lorentzian 
lines with different linewidths: a narrow and a broad one.  
Figure 2 presents the temperature dependence of the linewidths of the two signals.  
Similar two-component
EPR spectra were observed in other samples with different Sr
concentrations up to $x$= 0.06.  At $x$= 0.06, only a single
EPR line is seen in the entire temperature range, in agreement with our
previous studies of samples with 0.06 $\leq x\leq$ 0.20 \cite{Alika1}.
\begin{figure}[htb]
\includegraphics[width=1.0\linewidth]{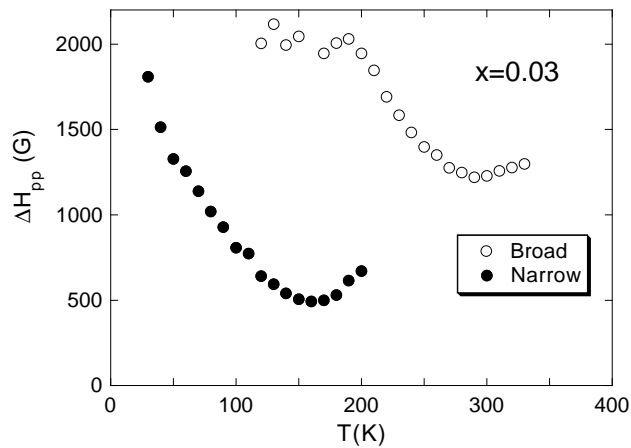}
\caption{Temperature dependence of the peak-to-peak linewidth 
$\Delta H_{pp}$ for the narrow and broad EPR lines in
La$_{1.97}$Sr$_{0.03}$Cu$_{0.98}$Mn$_{0.02}$O$_{4}$.}   
\label{Fig.2}
\end{figure}

Figure 3 shows the temperature dependence of the EPR intensity for
samples with different Sr concentrations.  One can see that the two
components observed for samples below $x$ = 0.06 follow a completely
different temperature dependence.  The intensity of the broad component 
strongly decreases with decreasing temperature.  On the other hand, the
intensity of the narrow component is negligible at high temperatures
and starts to increase substantially below $\sim$150 K. We note that the
temperature below which the intensity of the broad line decreases
shifts to lower temperatures with increasing doping.  However, the 
$I(T)$ dependence for the narrow line is practically doping-independent.  
It starts to increase almost exponentially below $\sim$150 K. 
A similar tendency is observed also for the temperature dependence
of the EPR linewidth.  The linewidth of the broad component and its temperature 
dependence are strongly doping-dependent, whereas the linewidth of the narrow 
component is very similar for all samples with different Sr doping.

\begin{figure}[htb]
\includegraphics[width=0.85\linewidth]{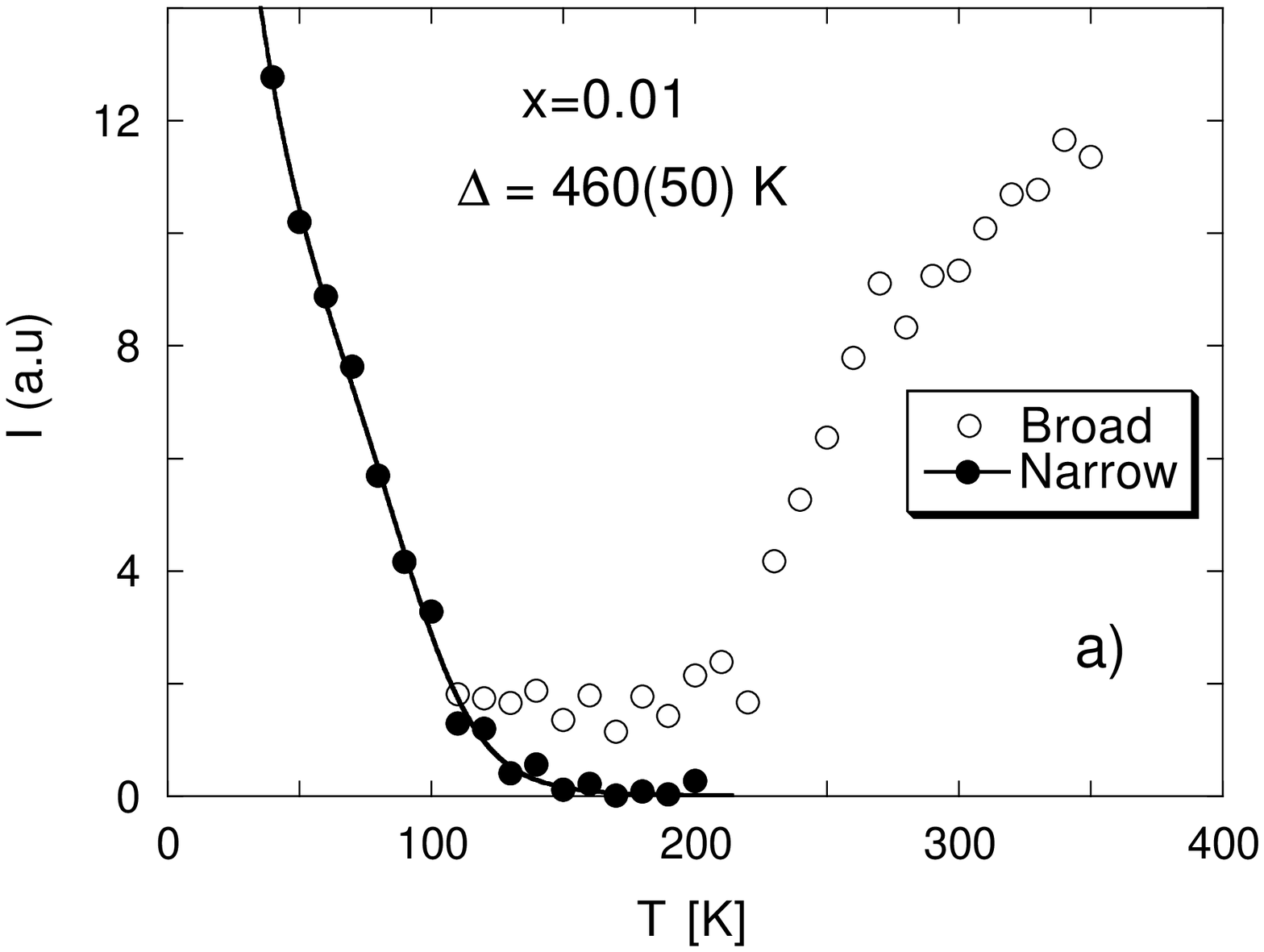}
\includegraphics[width=0.85\linewidth]{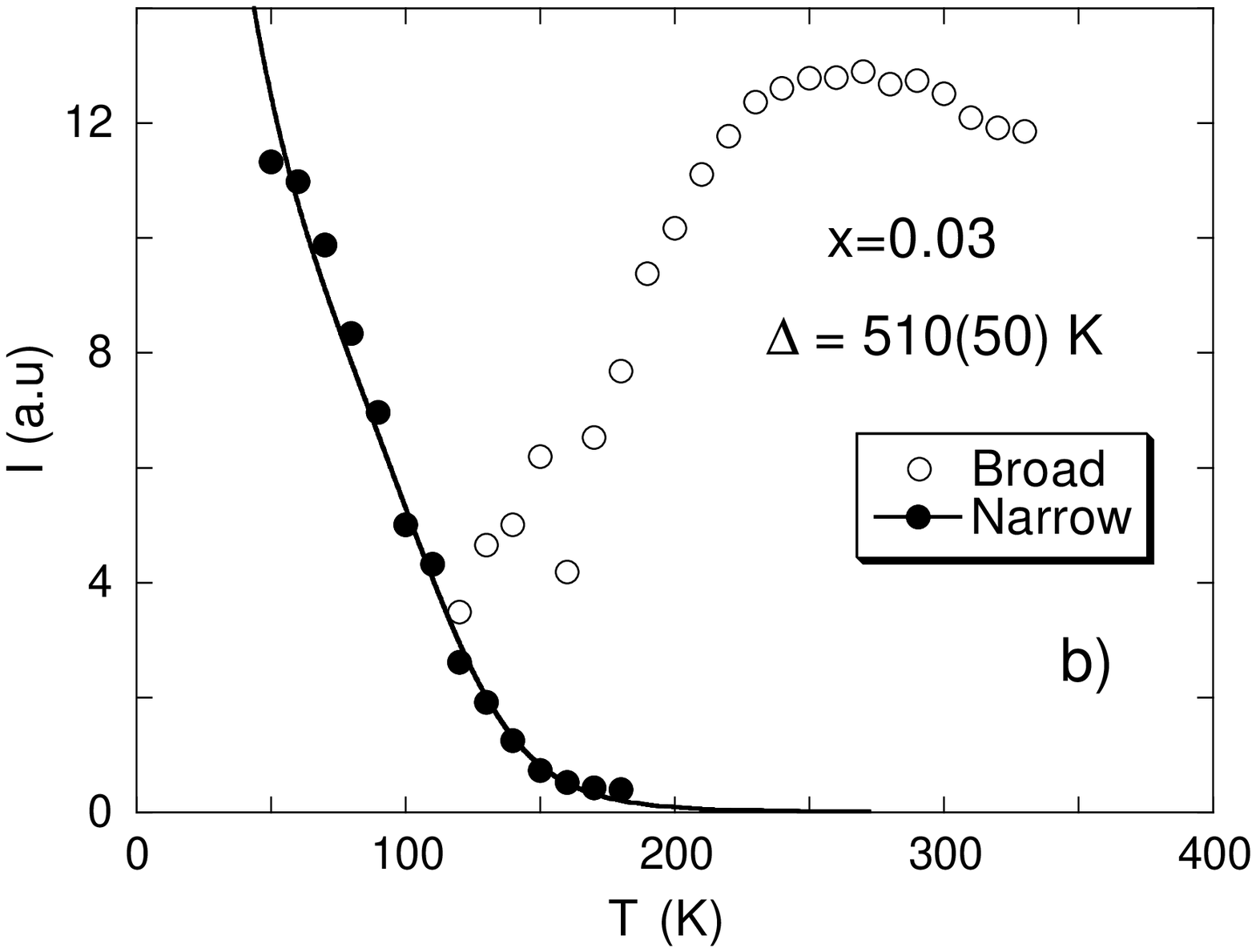}
\caption{Temperature dependence of the narrow and broad EPR signal 
intensity in  
La$_{2-x}$Sr$_{x}$Cu$_{0.98}$Mn$_{0.02}$O$_{4}$ with different Sr 
dopings: (a) $x$ = 0.01; (b) $x$ = 0.03. The solid lines 
represent fits using the model described in the text.}   
\label{Fig.3}
\end{figure}

It is important to point out that the observed two-component EPR spectra are 
an intrinsic property of the lightly doped LSCO and are not due to conventional 
chemical phase separation.  We examined our samples using x-ray diffraction, and 
detected no impurity phases.  Moreover, the temperature dependence of 
the relative intensities of the two EPR signals rules out macroscopic 
inhomogeneities and points towards a microscopic electronic phase separation.  
The qualitatively different behavior of the broad and narrow EPR
signals  indicates that they belong to distinct
regions in the sample.  First we notice that the broad line vanishes
at low temperatures.  This can be explained by taking into account the
AF order present in samples with very low Sr concentration
\cite{Johnston}.  It is expected that upon approaching the AF ordering
temperature, a strong shift of the resonance frequency and an increase
of the relaxation rate of the Cu spin system will occur.  This
will break the bottleneck regime of the Mn$^{2+}$ ions, and as a
consequence the EPR signal becomes unobservable \cite{Alika1}.

In contrast to the broad line, the narrow signal appears at low
temperatures and its intensity increases with decreasing temperature. 
This indicates that the narrow signal is due to the regions where the AF
order is supressed.  It is known that the AF order is destroyed by the
doped holes, and above $x$ = 0.06 AF fluctuations are much less pronounced
\cite{Niedermayer}.  Therefore, it is natural to relate the narrow line 
to regions with locally high carrier concentration and high mobility.  
This assumption
is strongly supported by the absence of an oxygen isotope effect on the
linewidth of the narrow line as well \cite{Alika2}.  It was shown 
previously that the isotope effect on the linewidth decreases at high
charge-carrier concentrations close to the optimum doping \cite{Alika1}.
We obtain another important indication from the temperature dependence of 
the EPR intensity. Because we relate the narrow 
line to hole-rich regions, an exponential increase of its intensity at low 
temperatures indicates an energy gap for the existence of these regions.
In the following we will argue that this phase separation is assisted by
the electron-phonon coupling.  More precisely, the latter induces
anisotropic interactions between the holes via the phonon exchange,
resulting in the creation of extended nano-scale hole-rich regions.

It was shown that the interaction between holes via the 
phonon exchange reduces to usual elastic forces if we neglect the 
retardation effects and optical modes \cite {Aminov}. In this case the
interaction between the holes  is highly
anisotropic, being attractive for some orientations and repulsive for
others \cite {Kochelaev}.  The attraction between holes may result in 
a bipolaron formation when holes approach each other closely enough.
The bipolaron formation can be a starting point for the creation of hole-rich 
regions by attracting of additional holes.  Because of the highly
anisotropic elastic forces these regions are expected to have the form
of stripes.  Therefore the bipolaron formation energy $\Delta$ can be
considered as an energy gap for the formation of hole-rich regions. 
In applying this model to the interpretation of our EPR results we have
to keep in mind that the spin dynamics of the coupled Mn-Cu system
experiences a strong bottleneck regime \cite{Elshner}.  In a
bottleneck regime, the EPR linewidth of the joint signal is controlled
mainly by the spin relaxation rate of the Cu spin-system, and the EPR
intensity is proportional to the sum of spin susceptibilities $I\sim
\chi_{Mn}+\chi_{Cu}$ \cite{Elshner}.  The latter results in a
Curie-Weiss temperature dependence of the EPR intensity as the spin
susceptibility of Mn is much larger than that of the strongly
correlated Cu spin system.

Taking into account this remark we conclude that the EPR intensity
of the narrow line is proportional to the volume of the sample occuppied
by the hole-rich regions because Mn ions are randomly distributed in the
sample.  We expect that in the underdoped samples the volume in
question is proportional to the number of bipolarons, which can be
estimated in a way used by Mihailovic and Kabanov \cite{Bipolaron}. 
If the density of states is determined by $N\left( E \right) \sim
E^\alpha$, the number of the bipolarons is

\begin{equation}
N_{\rm bipol} = \left( {\sqrt {z^2 + x} - z} \right)^2, \quad z =
KT^{\alpha + 1} e^{ - \frac{\Delta}{T}},
\end{equation}

\noindent
where $\Delta$ is the bipolaron formation energy, $x$ is the level
of hole doping, and $K$ is a temperature- and dopig-independent
parameter.  The EPR intensity from the hole-rich regions will be
proportional to the product of the Curie-Weiss susceptibility of the
bottlenecked Mn-Cu system and the number of the bipolarons

\begin{equation}
\label{inarl}
I_{\rm narrow}\sim \frac{C}{{T - \theta }}N_{\rm bipol}.
\end{equation}

The experimental points for the narrow-line intensity 
were fitted for the two-dimensional system 
($\alpha$=0), and we used the value $\theta$ = -8 K,
which was found from measurements of the static magnetic susuceptibility
(an attempt to vary $\theta$ yielded about the same value). 
The parameters $K$ and $\theta$ were kept the same for all samples;
the only free parameter was the energy gap $\Delta$. 
The results of the fit are shown in Fig.  3 (a,b).  For the bipolaron
formation energy we obtained $\Delta$=460(50) K, which is practically
doping-independent.  These values agree very well with the value of
$\Delta$ obtained from the analysis of inelastic neutron-scattering
and Raman data in cuprate superconductors \cite{Bipolaron}.
Recently Kochelaev et al.  performed theoretical calculations of the
polaron interactions via the phonon field using the extended Hubbard
model \cite{Kochelaev}.  They estimated the bipolaron formation energy and
obtained values of 100 K $\leq \Delta \leq$ 730 K, depending on the
value of the Coulomb repulsion between holes on neighboring copper and
oxygen sites $V_{pd}$, 0 $\leq V_{pd}\leq$ 1.2 eV. This means that
the experimental value of $\Delta$ can be understood in terms of the
elastic interactions between the polarons.

It is interesting to compare our results with other experiments
performed in lightly doped LSCO.  Recently
Ando {\em et al.} measured the in-plane anisotropy of the resistivity
$\rho_{b}/\rho_{a}$ in single crystals of LSCO with
$x$ = 0.02-0.04 \cite{Ando1}.  They found that at high temperatures the
anisotropy is small, which is consistent with the weak orthorhombicity
present. However, $\rho_{b}/\rho_{a}$ grows rapidly with decreasing 
temperature below $\sim$ 150 K. 
This provides macroscopic evidence that electrons
self-organize into an anisotropic state because there is no other
external source to cause the in-plane anisotropy in
La$_{2-x}$Sr$_{x}$CuO$_{4}$.  We notice that the temperature
dependence of the narrow EPR line intensity is very similar to that
of $\rho_{b}/\rho_{a}$ obtained by Ando {\em et al.} (see Fig. 2(d) in Ref. 4).  
To make this similarity clear, we plotted $I_{narrow}(T)$ and
$\rho_{b}/\rho_{a}(T)$ on the same graph (see Fig. 4).  It is
remarkable that both quantities show very similar temperature 
dependences. It means that our {\it microscopic} EPR measurements and 
the {\it macroscopic}
resistivity measurements by Ando {\em et al.} provide evidence of the
same phenomenon: the formation of hole-rich metallic stripes in lightly doped 
LSCO well below $x_{c1}$ = 0.06.  This conclusion is also supported by a
recent angle-resolved photoemission (ARPES) study of LSCO that
clearly demonstrated that metallic quasiparticles exist near the nodal
direction below $x$=0.06 \cite{Shen}.  

\begin{figure}[htb]
\includegraphics[width=1.0\linewidth]{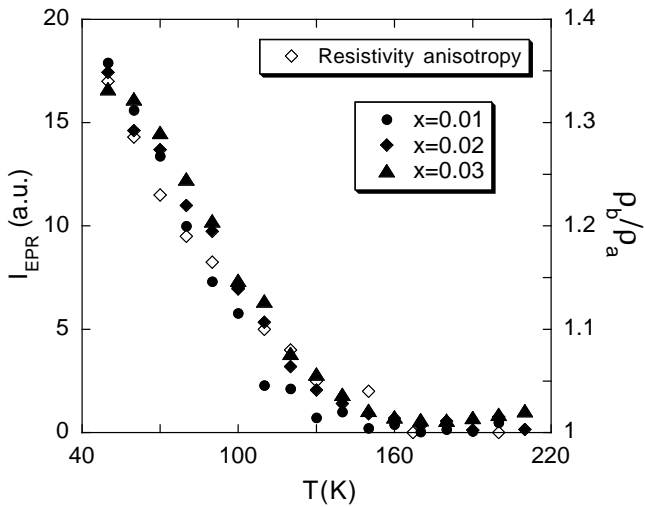}
\caption{Temperature dependence of the narrow EPR line intensities in
La$_{2-x}$Sr$_{x}$Cu$_{0.98}$Mn$_{0.02}$O$_{4}$ and 
of the resistivity anisotropic ratio in La$_{1.97}$Sr$_{0.03}$CuO$_{4}$
obtained in Ref.  4.}
\label{Fig.4}
\end{figure}

Finally, we would like to comment on the observability of the phase
separation in our EPR experiments.  The main difference of the EPR
signals from the hole-rich and hole-poor regions is the spin
relaxation rate of the Cu spin system, which results in different EPR
linewidths.  One would expect these local differences of the relaxation
rate to be averaged out by the spin diffusion.  The spin diffusion in
the CuO$_{2}$ plane is expected to be very fast because of the huge
exchange integral between the Cu ions.  A rough estimate shows that
during the Larmor period a local spin temperature can be transported
over 100 Cu-Cu distances.  It means that all the different nanoscale
regions will relax to the lattice with a single relaxation rate, and
we cannot distinguish them with EPR. However, the AF order which
appears below $T_N$ in the hole-poor regions in lightly doped LSCO
freezes the process of spin diffusion, and this is the reason we can
see different EPR lines from the two types of regions.  From this we
expect that with increasing doping, where magnetic order gets
suppressed, spin diffusion will become faster, extended, and we can no
longer distinguish different regions with EPR. This is most probably
what happens in samples with $x\geq$ 0.06, where only a single EPR
line is observed \cite{Alika1}.  This does not mean that the phase
separation in hole-rich and hole-poor regions does not exist at
$x\geq$ 0.06, but that the spin diffusion averages out the EPR
response from these regions.  In fact, recent Raman and infrared
measurements provided evidence of one-dimensional conductivity in LSCO
with $x$ = 0.10 \cite{Hackl}.  Also, recent ARPES measurements showed
that the nodal quasiparticle spectral weight changes smoothly with
doping without any anomaly at $x$ = 0.06 \cite{Shen}, indicating that
the electronic phase separation exists also at higher doping levels.

In summary, EPR measurements in lightly doped LSCO revealed the
presense of two resonance signals: a narrow and a broad one.  Their behavior
indicates that the narrow signal is due to hole-rich metallic regions
and the broad signal due to hole-poor AF regions.  The narrow-line intensity
is small at high temperatures and increases exponentially below $\sim$ 150
K. The activation energy inferred, $\Delta$ = 460(50) K, is nearly the
same as that deduced from other experiments for the formation of
bipolarons, pointing to the origin of the metallic stripes present. 
We found a remarkable similarity between the temperature dependences
of the narrow-line intensity and recently measured resistivity
anisotropy in CuO$_{2}$ planes in lightly doped LSCO \cite{Ando1}. 
The results obtained provide the first magnetic resonance evidence of
the formation of hole-rich metallic stripes in lightly doped LSCO well
below $x_{c1}$ = 0.06.

This work is supported by SNSF under the grant IP7-, BIK and AS are partially 
supported by 
INTAS-01-0654 and CRDF REC-007.


\begin{thebibliography}{99}

\bibitem{Tallon} J. F. Tallon {\em et al.}, Phys.  Rev.  B {\bf51},
12911 (1995).

\bibitem{Schneider}  See, e.g., T. Schneider, {\it The Physics of Conventional and 
Unconventional Superconductors}, edited by K. H. Bennemann and J. B. Ketterson 
(Springer Verlag, Berlin, 2003). 

\bibitem{Johnston} J. H. Cho {\em et al.}, Phys. Rev. Lett. {\bf70}, 222 (1993).

\bibitem{Ando1} Y. Ando {\em et al.}, Phys. Rev. Lett. {\bf88}, 137005 (2002).
   
\bibitem{Alika1} A. Shengelaya {\em et al.}, Phys. Rev. B {\bf63}, 144513 (2001).

\bibitem{Alika2} A. Shengelaya {\em et al.}, J. Supercond, {\bf13}, 955 (2000).

\bibitem{Bipolaron} V. V. Kabanov and D. Mihailovic, Phys.  Rev. 
B{\bf 65}, 212508 (2002).

\bibitem{Niedermayer} Ch. Niedermayer {\em et al.}, Phys. Rev. Lett.  {\bf80}, 3843 (1998). 

\bibitem{Aminov} L. K. Aminov and B. I. Kochelaev, Zh. Eksp. Theor. Fiz. {\bf 42}, 
1303 (1962).

\bibitem{Kochelaev} B. I. Kochelaev {\em et al.}, Mod. Phys. Lett.  B
{\bf17}, 415 (2003).

\bibitem{Elshner} B. I. Kochelaev {\em et al.}, Phys. Rev. B {\bf 49}, 13106 (1994).

\bibitem{Shen} T. Yoshida {\em et al.},  Phys. Rev. Lett. {\bf 91}, 027001 (2003).

 \bibitem{Hackl} F. Venturini {\em et al.}, Phys. Rev. B {\bf66}, 
 R060502 (2002).

\end{thebibliography}
\end{document}